# Future Network: End-to-End Slicing and Hop-On (a Slice)

Hang Zhang, Huawei Technologies Canada, Ottawa, Ontario, Canada,
E-mail: hang.zhang@huawei.com

The concept of network slice (a service customized virtual network –VN) is attracting more and more attentions in the telecommunication industry. A slice is a set of network resources which fits the service attributes and requirements of customer services. The network resources consist of cloud resources and communication link resources. A slice can serve one or more customer services which share the similar service attributes and requirements. To define, create and manage a slice (VN) is one aspect of future networks. Another aspect is the slice operation, i.e., the provisioning of services to customers using created slices. In this paper, the focus is put on the configuration of slices and the operation of slices. In this paper, the detailed description of slice (VN) configuration is provided. A new concept of hop-on (a slice) is described. Given a well defined and configured end-to-end slice (VN), the real-time data traffic delivery over a slice is governed by network operation control entities, which are also pre-configured. Therefore, the procedure of customer traffic delivery over a slice is just like a traveler hopping on tourist bus and then the traffic control officers at key intersections directing the traveler to go through pre-designed routes until the destination is reached.

## 1. Introduction

Network slice (or a service customized virtual network –VN) is attracting more and more attention in the telecommunication industry. There are many definitions of slice. NGMN [1, 2] and 3GPP [3] have contributed a lot on slice definition. In short, a slice is a set of network resource which fits the service attributes and requirements of customer services. The network resource includes cloud resource and communication link resource. A slice can serve one or more customer services which share the similar service attributes and requirements. Customer services here include vertical services (e.g., industry control), over-the-top (OTT) customer services, operator defined services for a group of subscribers, etc. Such a service usually involves many devices.

Hop-On (a slice) means that, given a well defined slice, any of end-points (devices and servers) of a slice should be able to directly use the pre-defined slice to transmit data traffic without per-device or per session triggered end-to-end connation establishment. An analogy can illustrate this concept. Travelers are travelling from a variety of origins to a variety of destinations. In order to optimize the transportation resource utilization and to best fit the traveler's requirement, a well defined tourist transportation system is designed. During the operation of this system, traffic control officers at key intersections direct travelers onto the right pre-defined routes. Here the transportation system represents the pre-configured slice (VN) and the traffic control officers at key intersections represent pre-configured control entities.

To make human-free slice definition, creation and adaptation, MyNET network architecture and SONAC techniques are briefly discussed in [4]. SONAC techniques include two parts: SONAC for slice composition, denoted as SONAC-Com; SONAC for slice operation, denoted as SONAC-Op. SONAC-Com defines and creates slices and modifies slices to adapt to various changes of statues. SONAC-Op controls and manages the connectivity service provisioning to end-points using a pre-defined slice. SONAC-Com controls and manages resource for a service at slice level. SONAC-Op controls and manages resource for devices during slice operation.

SONAC-Com consists of three components: SDT-Com (software-defined-topology), SDRA-Com (software defined resource assignment), and SDP-Com (software defined protocol). Similarly, SONAC-Op consists of three components: SDT-Op for device routing control over a VN during the operation; SDRA-Op for device physical resource assignment; SDP-Op for protocol selection for a device/application.

Three key enabling techniques for Hop-On (a slice) are SONAC-Com, SONAC-Op and Connectivity management (CM) [4].

Techniques of network function virtualization provided by network function virtualization (NFV) [5, 6, 7] and software defined networking (SDN) [8, 9, 10] have laid the foundation of future network design. In this paper, the availability of network programmability provided by NFV and SDN is assumed.

In the following discussion, for the sake of easy description, we assume that a slice serves one customer service and use VN (virtual network) to denote a slice. The concept described here, however, is applicable to the cases where a slice serves multiple customer services with minor extension.

This paper is organized as follows:

- Background and terminology
- VN configuration - How to realize a slice in a network infrastructure, with focusing on the slice configuration by SONAC-Com
- VN operation – How to operate a VN by SONAC-Op
- Hop-On (a slice) – How to deliver service to end-points over a pre-defined VN (slice)
- Benefits of slicing and hopping-on a slice
- Summary

## 2. Background and terminology

For an easy description, a small network infrastructure is used in this paper, as shown in Figure 1. For location tracking purpose, a network can be divided into domains. Each domain could be further divided into clusters. In this figure, two layers are assumed. However, there could be more or less layers and the division of a network is completely determined by operators.

In this figure, there are two domains and there are two RAN clusters within each domain. There are multiple physical network nodes (NNs) and multiple pieces of cloud. Hierarchical SONAC architecture and hierarchical CM architecture are shown. The CM entities at different layers and the inter-connections among them can be viewed as a CM slice. The logical interconnections between SONAC entities at different layers of the architecture are not shown for the sake of simplicity. For the same reason, interconnections between CM entities at the different layer are not shown either.

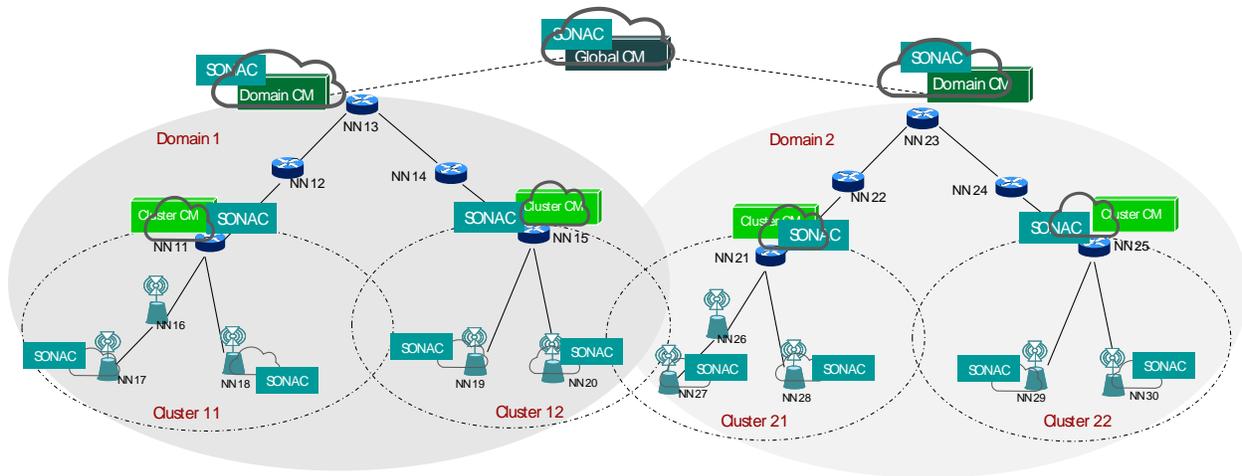

Figure 1. Example of wireless network infrastructure topology.

A RAN cluster consists of one or multiple access nodes, which are interconnected via wired or wireless links. Access nodes within a RAN cluster are general nodes that could be an access node with RF components and all, part or no baseband process.

SONAC-Com functions and CM functions can be pre-realized before any customer service slice is defined and implemented. SONAC-Op can be pre-realized if it controls and manages all service slice operations. SONAC-Op can also be created along together with a service slice if SONAC-Op is dedicated to the slice.

A slice is a virtual network (VN) which is set up using a set of physical network infrastructure resource.

Such a VN is composed of a number of VN Nodes. A VN Node is defined as a virtual VN node which locates in a cloud and associates with a physical NN. A VN Node could consist of a network function (NF) or a group of NFs of a VN. Logical topology of a VN is the inter-connection among these VN nodes which are distributed in the selected locations for the VN. A VN tunnel (or a slice tunnel or a service tunnel) is defined as the logical connection between two of VN Nodes of a VN. In a RAN cluster, one open VN tunnel is defined as the logical connection between a VN Node and an access node where there is no a VN specific NF located. End-points of a VN are devices and application servers that use the VN resource to obtain connectivity service.

3G/4G wireless networks have successful provided connectivity services to millions of millions of mobile subscribers in the past decades. As the increase of the demands of customers on wireless network services, the 3G/4G networks is facing challenges to deal with the billions of billions of machines/devices communication (connection), services with supper high reliability and low latency requirement, and so on. We will discuss the how the concepts of slice and hopping –on a slice help operators to deal with these challenges.

## 3. VN configuration

SONAC-Com needs to configuration of a VN after the definition and NFs instantiation. The functionalities of SDT-Com and SDRA-Com for configuring a VN and configuring SONAC-Op are discussed in this section.

### 3.1 SDT-Com functionality

The main tasks of SDT-Com is to define a VN, i.e., determines the NFs needed for a slice, the positions where NFs should be implemented in a network infrastructure, and the topology of a VN. After this, SDT-Com needs to provide the decision to SDRA-Com, SDP-Com and the entity which instantiates NFs in cloud (ETSI NFV defined VIM (virtual infrastructure manager) will be the entity [5, 6, 7]). In this paper, we will focus on the configuration of VN after a slice has been defined, the logical topology has been designed and necessary NFs have been instantiated. In addition, SDT-Com also needs to configure SDT-Op for guiding the data packets routing over a VN during the VN operation.

*VN description:*

- VN Node description:
    - VN node ID: Physical NN ID to which a VN node associates, ID of domain or cluster where the VN Node is the anchor point of a VN in this domain or cluster
- VN logical topology description
    - Tunnel ID: ingress VN Node ID, Egress Node ID, QoS
- VN open logical topology description
    - Open tunnel ID: ingress VN node ID (source NN ID), destination NN ID (VN Node ID)
- End-point QoS
    - Rate, Latency, etc

*Configuration of SDT-Op (VN routers) for a created VN:*

- Association of VN Node to a VN router
    - VN Node ID: VN router ID
- Tunnel configuration
    - Tunnel ID: egress VN node ID, QoS
- Open tunnel configuration
    - Open tunnel ID: egress VN Node ID (destination NN ID)
- Configuration of VN routing table:
    - Destination VN Node ID: next VN Node ID

A VN router also needs to associate with connectivity management (CM) entity for the purpose of location information acquisition.

- Associate a VN router with a CM entity
    - VN router ID: CM ID

Figure 2 shows one example of VN set up using resource in domain 1 of the network infrastructure shown in Figure 1. Figure 2 (a) shows VN (VN ID = 1) description. Figure 2 (b) shows the tunnel configuration of SDT-Op (VN router) and Figure 2 (c) shows the configuration of VN routing table of VN routers.

***Discussion of tunnel QoS definition:***

Topology design of a VN is based on statistics of data traffic attributes, geographic distribution of end-points and quality requirement. To determine QoS of a VN (slice) tunnel, SONAC-Com use statistical models to estimate the level of traffic that will go through the tunnel.

This will be used for VN admission control and used by SDA-Op for device traffic physical link resource assignment on this tunnel.

Similarly, for a RAN cluster, the QoS of a tunnel or open tunnel are also defined based on statistics of data traffic. In some cases, the tunnel definition doesn't explicitly indicate QoS for each such a tunnel to enable SDRA-Op perform real-time resource management to maximize resource utilization.

In addition to the rate definition, packet latency budget of a tunnel may need to be explicitly indicated by QoS definition for some services. When SDRA-Com maps a logical tunnel to physical network resource, the packet latency budget of a logical tunnel will be used for calculating the latency budget allocated to each of physical links.

For a service, there may be more than one type of packets with each of these types requiring different treatments, such as control packets and data packets of a smart reading service. To differentiate these packets of a single service, a packet type ID or priority header can be used.

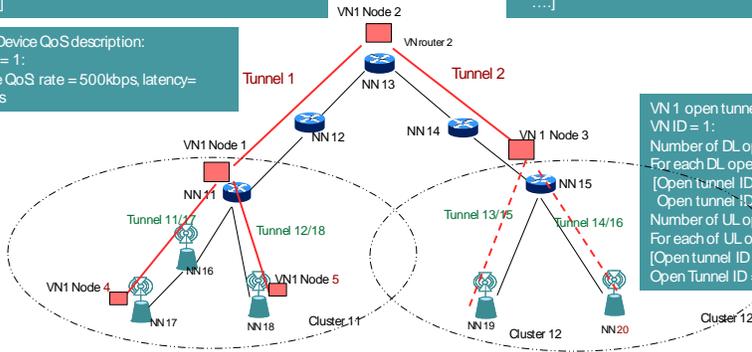

(a) VN description.

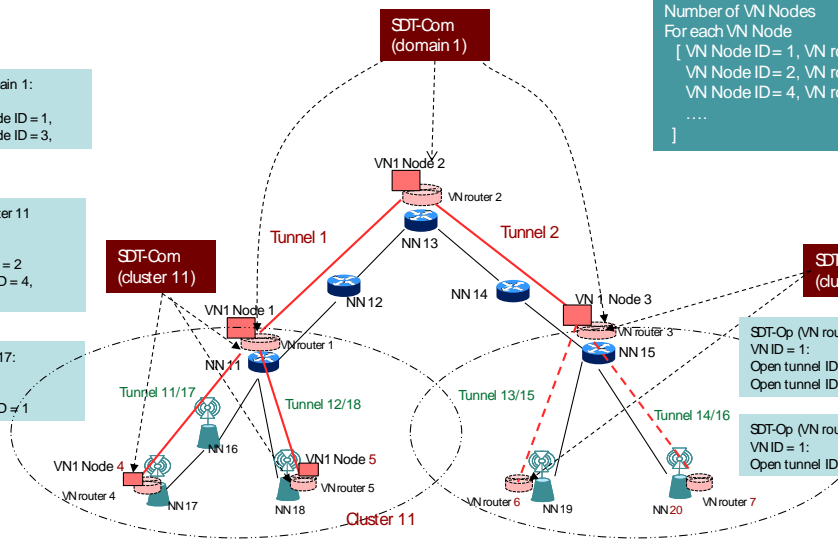

(b) Tunnel configuration of VN router.

- SDT-Op: Implemented by VN routers
- Routing table configuration

VN router 2
VN ID=1:
Destination VN Node ID = 4, tunnel ID = 1;
Destination VN Node ID = 5, tunnel ID = 1;
Destination VN Node ID = 1, tunnel ID = 1;
Destination VN Node ID = 3, tunnel ID = 2;
]

VN router 1
VN ID=1:
Destination VN Node ID = 4, tunnel ID = 11;
Destination VN Node ID = 5, tunnel ID = 12;
…]

VN router 4
VN ID = 1:
All destination VN Nodes , tunnel ID = 17;

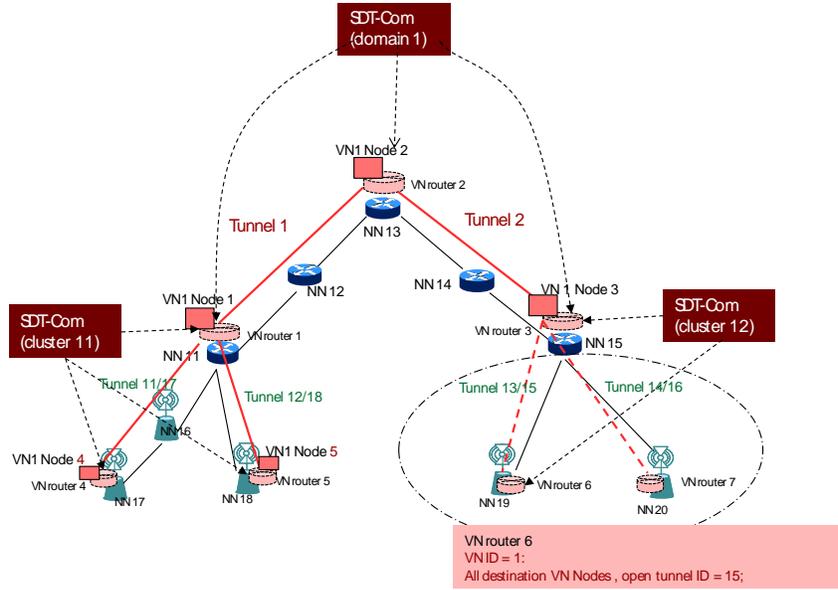

(c) VN routing table configuration of VN router.

NN 11: VN ID = 1;
Destination NN ID = 17 , next NN ID = 16, QoS
Destination NN ID = 18, next NN ID = 18, QoS
…]

NN 16: VN ID = 1;
Destination NN ID = 17, next NN ID = 17, QoS
Destination NN ID = 11, next NN ID = 11, QoS

NN 17: VN ID = 1
Destination NN ID = 11, next NN ID = 16, QoS;
**Access Link (AL):**
DL: VN ID = 1: QoS (rate, etc) (per service/per device),
AL resource pre-assignment = YES, resource ID;
Service based (or device based) resource = YES;
UL: VN ID = 1, QoS (rate, etc) (per service/per device),
AL resource pre-assignment = YES, resource ID;
Service(device) based resource = YES;

NN 18:
Destination NN ID = 11, next NN ID = 11, QoS;
**Access Link (AL):**
DL: VN ID = 1: QoS (rate, etc) (per service/per device),
AL resource pre-assignment = NO;
Service based (or device based) resource = NO;
UL: VN ID = 1, QoS (rate, etc) (per service/per device),
AL resource pre-assignment = NO;
Service(device) based resource = NO;

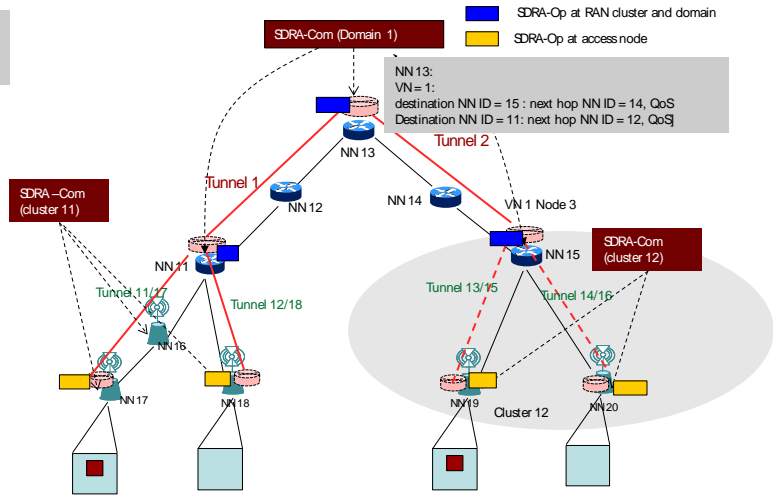

(d) Configuration of mapping between logical tunnel to physical network resource.

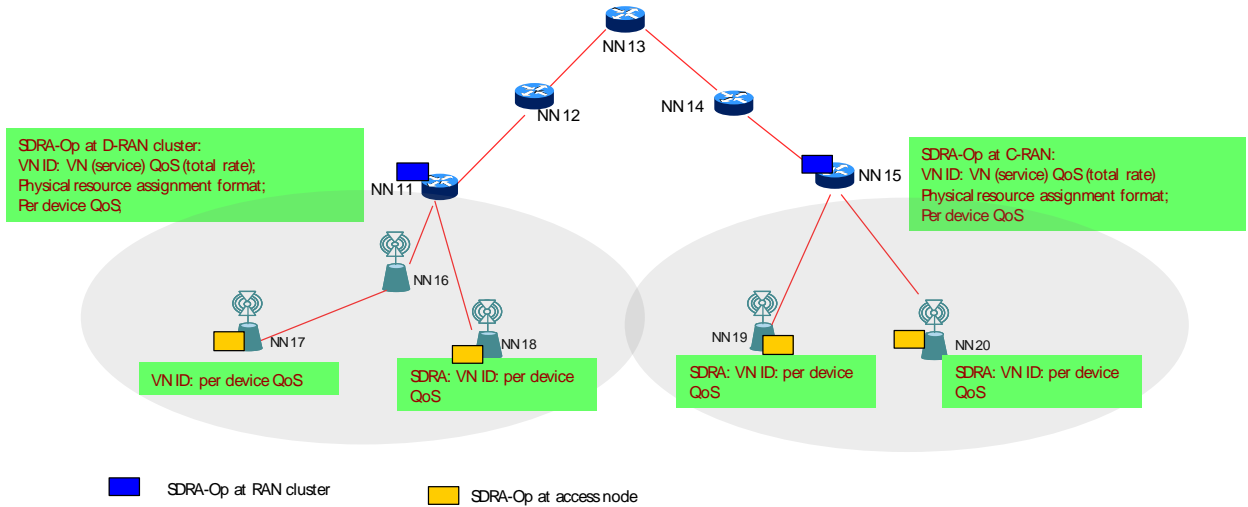

(e) Configuration of SDRA-Op.

Figure 2. VN configuration by SONAC-Com.

At his point, a logical VN is completely configured and the required configuration of SDT-Op has been completed. Next step is to map the logical tunnel into physical network resource.

## 3.2 SDRA-Com functionality

SDRA-Com is the component of SONAC-Com for mapping logical tunnels to physical network resource. SDRA-Com also needs to configure SDRA-Op.

***Mapping between a logical tunnel of a VN to physical network resource:***

SDRA-Com performs logical tunnel mapping to physical network resource. For each logical tunnel, there could be multiple physical paths and multiple hops. For each logical tunnel of a VN, SDRA-Com needs to determine the mapping format and need to configure the involved NNs.

There are multiple formats of this mapping:

- For IP- like routing
    - Configure NNs the format= IP-like routing
- Source routing
    - Configure NNs the format = source routing
- Destination based routing
    - Configure NNs the format = destination
    - Configure forwarding rule to NN:
        - VN ID (service ID): destination ID, QoS, next hop NN ID
- Dedicated resource
    - Configure NNs the format = dedicated resource

- Configure forwarding rule to NN:
  - VN ID (service ID): dedicated resource (lambda, or sub-spectrum band, etc)

Only a few mapping formats are listed here.

For access link, the formats include
- Dedicated resource, i.e., dedicated air interface resource (in time, frequency and code domains)
  - Configure NN the format = dedicated AL resource
    - VN ID, resource assignment (ID)
- Shared resource
  - Configure NN the format = shared AL resource, i.e., there is no dedicated AL resource is assigned

The mapping format of dedicated resource is used for VN which requires guaranteed low latency. For all other formats, the mapping is only "logical", i.e., the mapping is only to instruct physical NN on how to treat received packets of a VN.

***Configuration of SDRA-Op:***

During the operation, SDRA-Op needs resource assignment to data packets of devices.

For a RAN cluster, the integrated rate requirement at RAN cluster (service level) is estimated and indicated to SDRA-Op at that RAN cluster. SDRA-Op will use this information for resource assignment of devices to make sure that the service level QoS can be satisfied and, at the same time, there is no over-provisioning.

For access link, per device service requirement should be indicated to access nodes such that access link SDRA-Op at access node (AL scheduler) can make decision on resource assignment to devices.

At this point a VN with physical configurations are completed. Please note that the SDP-Com takes care of the end-point protocol configuration and tunnel protocol configuration, which is not the focus of this paper.

Figure 2 (d) and (e) show a completely defined VN with mapping of logical tunnel to physical network resource, and the SDRA-Op configuration, respectively.

At this point, the VN is ready to go.

# 4. VN operation

After a VN is completely configured, SONAC-Op govern the slice operation together with other network operation functions (e.g., CM).

## 4.1 Device Registration and hop-on (a slice)

*Registration:* If an end-point (device or server) wants a service that can be provided by a VN, the device should register to this VN either instructed by network or by the end-point with pre-configured policy. There are many different ways to perform the registration. After the registration, a device should obtain VN ID (or service ID), and radio ID (similar to MAC ID in 3G/4G) used for AL resource assignment. From network side, after this procedure, a device can be associated with a VN Node. It means that this VN node will be an anchor point of the end-point in a VN. The association of a device to a VN Node will be determined based on device MAC state and mobility velocity and so on. This association between a device and a VN node can be viewed as logical tunnel between the device and the VN Node. A device registration procedure is highlighted in Figure 3.

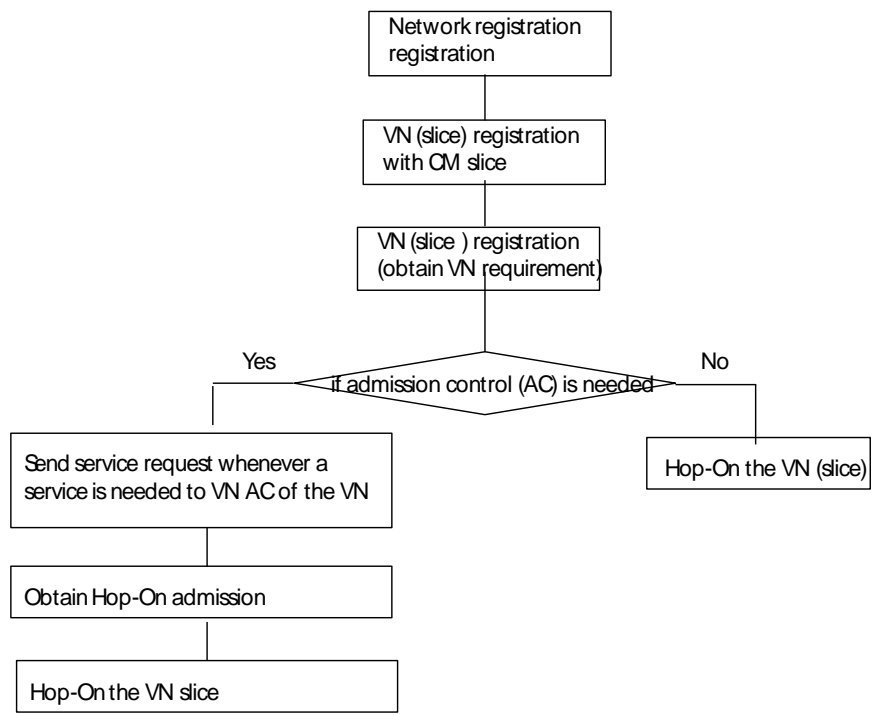

Figure 3. End-point operation procedure for hopping on a VN.

- Network registration: get authentication ad authorization to use the network
- Registration to a CM: start reach-bility operation to enable network track the location of the device even before register to any user plane (UP) VN (slice).
- Register to a UP slice: get authorization to use a UP slice and obtain information on the requirement whether an admission control (AC) is needed for follow up service data transmission
- Service admission control
    - If No, devices/servers can send data over the slice at any time without exceed the assigned rate

- If Yes, whenever a device/server needs a service (e.g., download file with certain rate, etc), send a request and start hop-on after get admission

*Hop-On*: For hopping on a VN, device simply sends data packets which each of them carrying VN ID (or service ID) and the name or ID of the destination. Based on the VN ID, network will forward the data tunnel by tunnel over a VN until the data packet reaches the destination. The VN routers associated with VN Nodes of the VN, needs to decide the next tunnel ID (or next VN node ID) after a data packet is received and processed in a VN Node. This process needs an end-point routing table at each VN router. For fixed end-points, the table can be obtained after end-point initial location registration. For mobile devices, a VN router needs to acquire the location information from its associated CM entities. Based on the destination name (or ID) and the acquired location information, a VN router will route the data packet to the right next tunnel.

## 4.2 Operation of SDT-Op

*VN end-point routing table and information:*

Interaction between VN router and CM is shown in Figure 4 to illustrate the device location tracking and location resolution operation. Name (or ID) based location tracking using the hierarchical CM architecture shown in Figure 1 is used in Figure 4.

Simplified figure in Figure 4 (a) is used to highlight the name based location tracking of a device. The interaction and location resolution procedure are illustrated in Figure 4 (b).

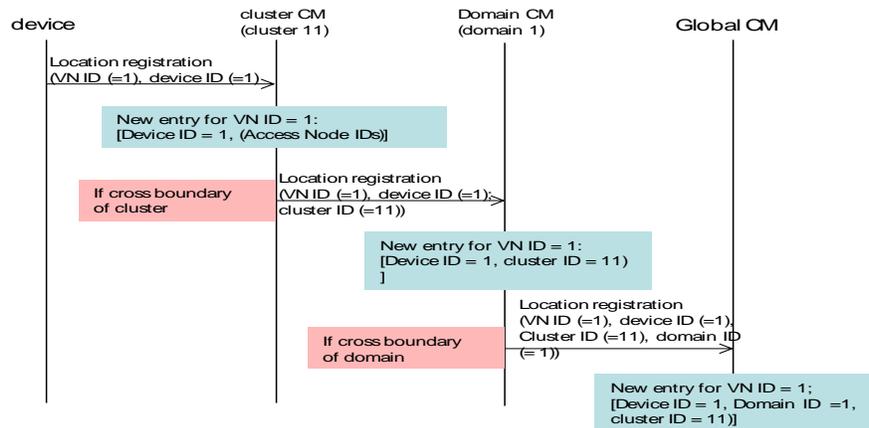

(a) Location tracking procedure.

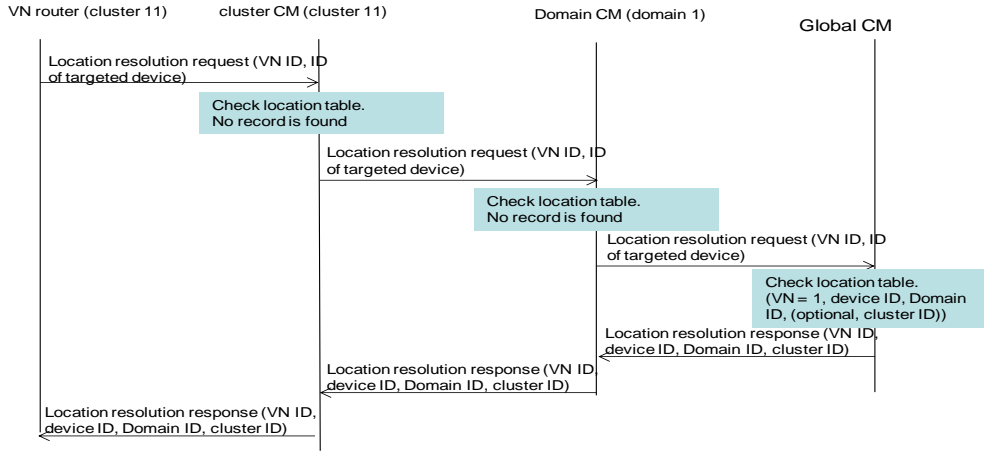

(b) Location resolution procedure and interaction between SDT-Op and CM.

Figure 4. Location tracking and location resolution of mobile devices by CM.

The location registration procedure shown is an example with device explicit location registration. This procedure can be implicit. A device may only report DL measurement reports to cluster CM or only send UL sequence. For latter case, access nodes which are configured to monitor this sequence report the measurement strength to the cluster CM. A cluster CM always maintains the information of candidate set of access nodes (tunnels) for a device which actively receiving data from network.

For location resolution, there are many approaches to reduce the involved signaling. The returned location of a device can be at domain level, cluster level or access node level. This can be configured by CM. The location resolution scheme can be requesting or pushing mode, and can be different at different level of the CM hierarchy.

***Operation of SDT-Op (VN router) for fixed end-points of a VN:***

After the location registration, the VN routers should create an end-point routing table which includes information, such as, VN ID and for each end-point ID, next VN node ID, as shown in Figure 5 (a).

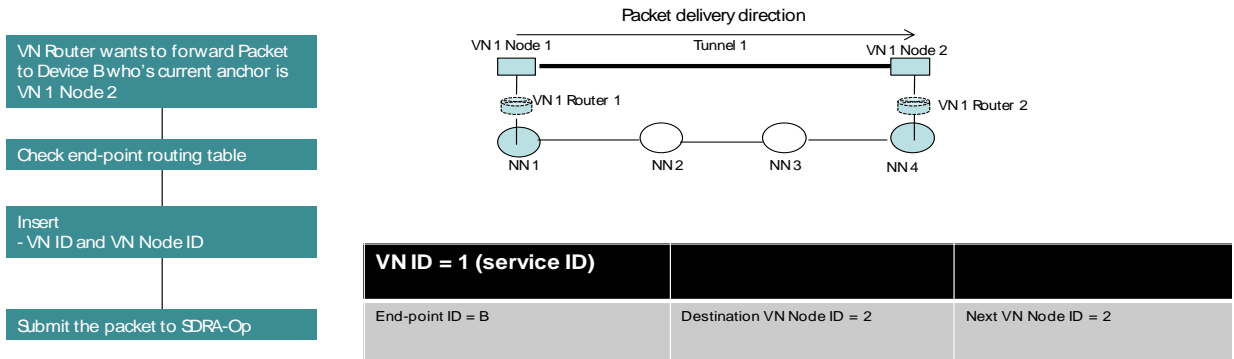

(a) VN serving fixed end-points.

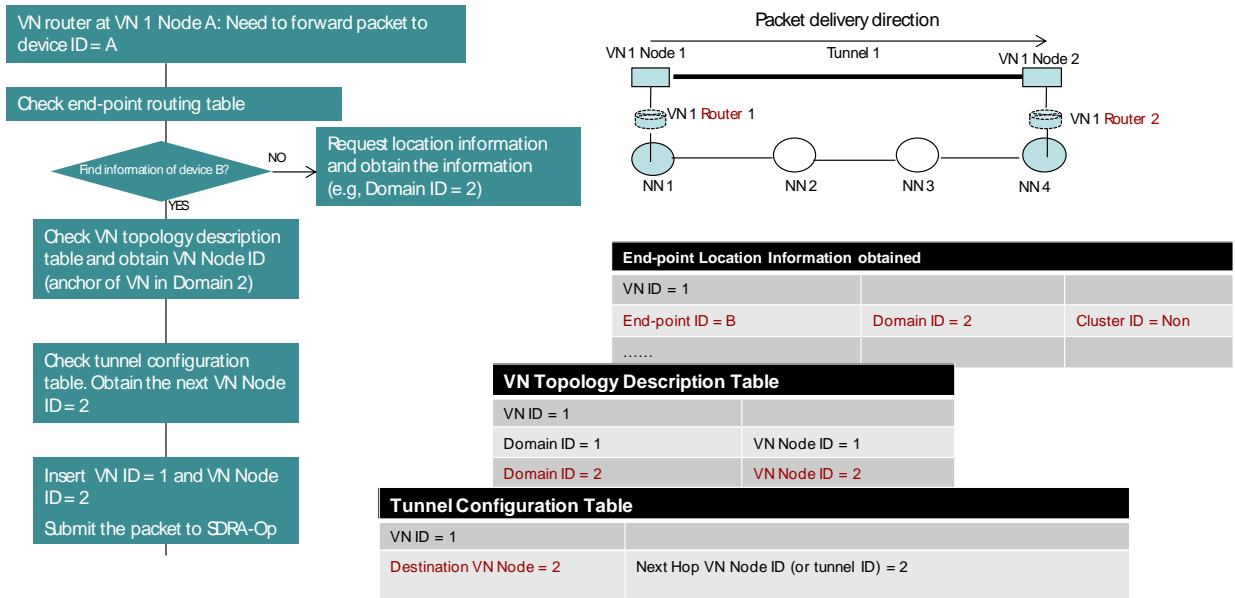

(b) VN serving mobile devices.

Figure 5. Operation of SDT-Op (VN router).

In this example, VN router 1 needs to forward data packet to destination end-point device B which is associated with VN Node 2. VN router 1 checks the end-point routing table and knows that the next VN node ID = 2. The VN router inserts VN ID and VN node ID to the packet and submits it to SDRA-Op.

*Operation at VN router for mobile end-points of a VN:*

For a VN which provides services to mobile devices, the VN router needs to obtain current anchor VN Node of this mobile device or domain ID where the mobile device is currently in. The procedure of a VN router routing data packet to a mobile device is shown in Figure 5 (b). For data traffic routing to a mobile device, the VN router needs to obtain the location information (domain ID or cluster ID) from CM to which the VN router associates. VN router needs to check the VN description table and translate the location information to the destination VN Node ID. After checking VN routing table, VN determines the next VN Node ID or Tunnel ID. Then the VN router submits the data packet with VN ID and VN Node ID to SDRA-Op.

Same as the case for fixed devices, a VN router needs to keep an end-point routing table for mobile devices after getting location information from CM. The rule to update the table needs to be configured to make the VN routing procedure efficient.

In RAN clusters, a VN with multiple tunnels or multiple open tunnels are defined and realized to enable handover-free to mobile device or to support customer services with high reliability requirement.

At RAN cluster, to deliver a data packet to a mobile device, the VN router in this cluster needs to check with the CM to obtain the location information (candidate access nodes of a mobile device). The VN router then needs to determine the destination access nodes (or tunnels) and submit the packet with multiple destination node IDs (multipath – different data packets to different destination access nodes, or multicast – same data packet to all destination nodes) and VN ID to SDRA-Op of this cluster.

## 4.3 Operation of SDRA-Op

***SDRA-Op operation at ingress of a tunnel (packet header process):***

After receiving a packet from VN router, based on the VN ID and VN Node ID (or tunnel ID), the SDRA-Op checks the physical resource mapping format and the resource assigned, process the packet header as below:

- For IP-like routing
    - Change VN Node ID to IP address of the NN to which the VN Node associates, Insert QoS field
- For source routing
    - Change VN Node ID to source routing header, insert QoS field
- For label based routing format:
    - Remove VN Node ID (keep VN ID)
- For dedicated resource
    - Remove VN Node ID (keep VN ID)

After this procedure, the packet is submitted to physical network node (NN) and the data packet is forwarded along the physical path until it reaches the egress of the logical tunnel.

A logical tunnel of a VN can be mapped to multiple physical paths. For this case, for each received packets from VN router, SDRA-Op needs to determine the rate splitting of data flow over these physical paths and the source routing is the easiest way to support this.

***SDRA-Op at RAC cluster:***

At RAN cluster for DL data packets to a mobile device, a VN router needs to determine the access nodes (i.e., tunnels) to which the data packet can be forwarded. For multiple destination access nodes (tunnels) and multipath case, SDRA-Op need to decide the rate splitting over those tunnels of a VN with the consideration of physical resource multiplexing of multiple VNs.

***SDRA-Op at access link:***

At access node, for VN with pre-assigned AL resource, the scheduler assigns resource to individual device based on the per-device QoS within the pre-assigned VN resource in this AL. For VNs without pre-assigned AL resource, the AL scheduler assigns resource to devices based on the per-device QoS and per-service QoS configured during the VN realization.

Figure 6 (a) shows the general actions of SDRA-Op and (b) shows the process of SDRA-Op at a RAN cluster and at an access node.

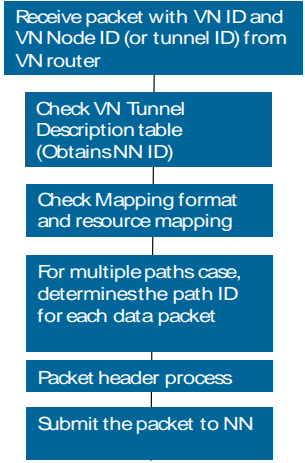
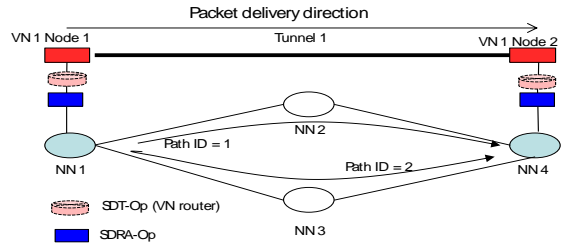

(a) General process of SDRA-Op.

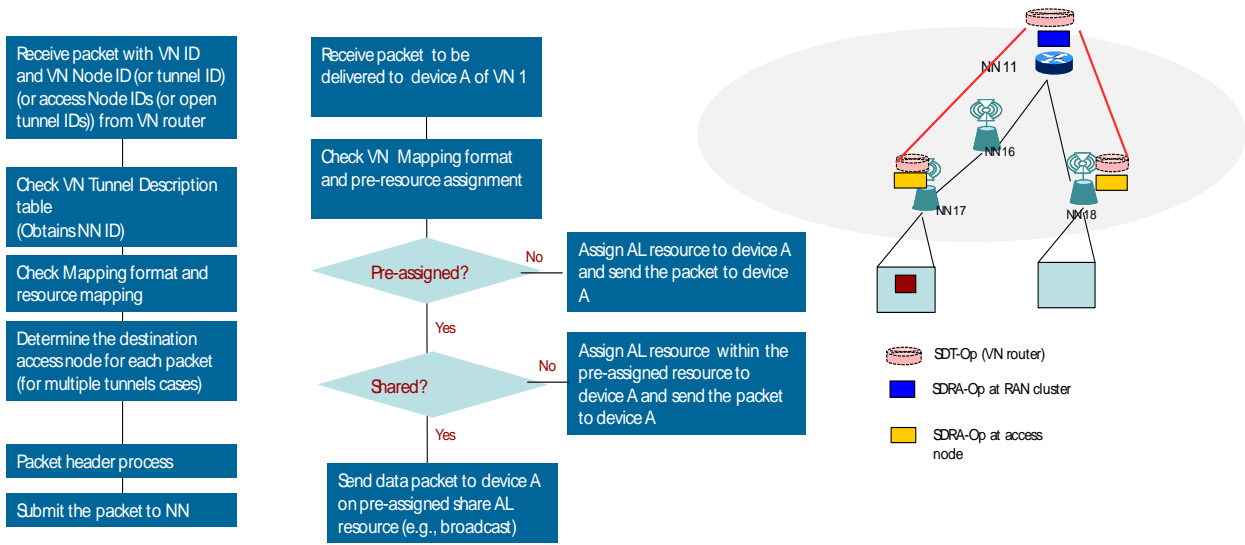

(b) SDRA-Op at RAN cluster and access node.

**Figure 6. Operation of SDRA-Op.**

## 4.4 Interaction between SDT-Op and SDRA-Op

The interaction between SDT-Op and SDRA-Op is shown in Figure 7. In this figure, it is assumed that SDT-Op and SDRA-Op provide service to multiple VNs. In most of cases, SDT-Op receives packets from application functions, process the packet (determines the next tunnel) and sends the packet directly to SDRA-Op. In some other cases, SDT-Op sends the processed packet to other NF first (e.g., packet aggregator) and then the NF send the packet to SDRA-Op. This can be configured by SONAC-Com during VN configuration phase. SDT-Op and SDRA-Op can also be per VN specific.

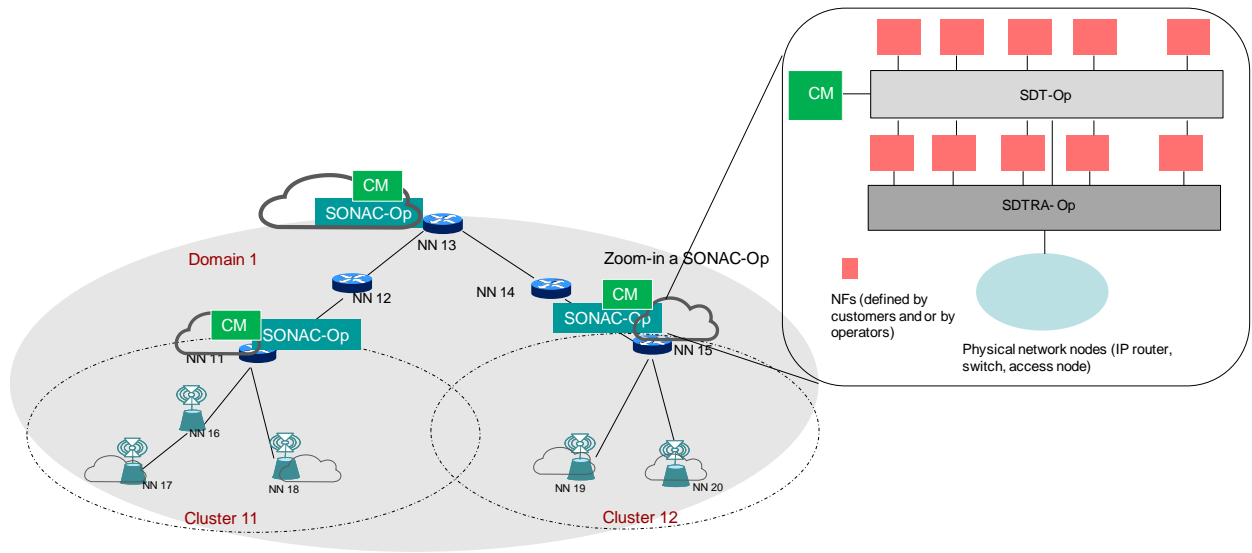

Figure 7. Interaction between SDT-Op and SDRA-Op.

# 5. Hop-On a VN (Slice)

What is Hop-On? Since the VN topology and routing table and the mapping between logical tunnels to physical network resource have been pre-configured, no and little signaling is needed for data traffic delivery for most of use cases. End-point simply sends packet with the VN ID (service ID) and the name (ID) of destination end-point to the network after obtaining the authorization of using a VN. VN routers perform the routing of packets using the location information. To enable this, the location of an end-point needs to be tracked and the location information needs to be accessible to VN routers. Sending devices don't need to worry about the location of the destination devices. Data packets of end-points go over VN along the route guided by SDT-Op (VN routers).

The end-to-end packet routing procedures for fixed end-points and for mobile end-points are shown in Figure 8 (a) and (b), respectively. For the sake of simplicity, it is assumed that the pre-assigned share AL resource (i.e., UL contention resource and DL broadcast resource are assigned to a VN) are configured. Figure 8 (b), the location information is retrieved by the first VN router (e.g., at an access node). However, the location information can also be retrieved by other VN router, e.g., a VN router at a RAN cluster. For this case, VN router at the access node will route packets to VN Node at the RAN cluster from where the location is retrieved by the VN router at that RAN cluster.

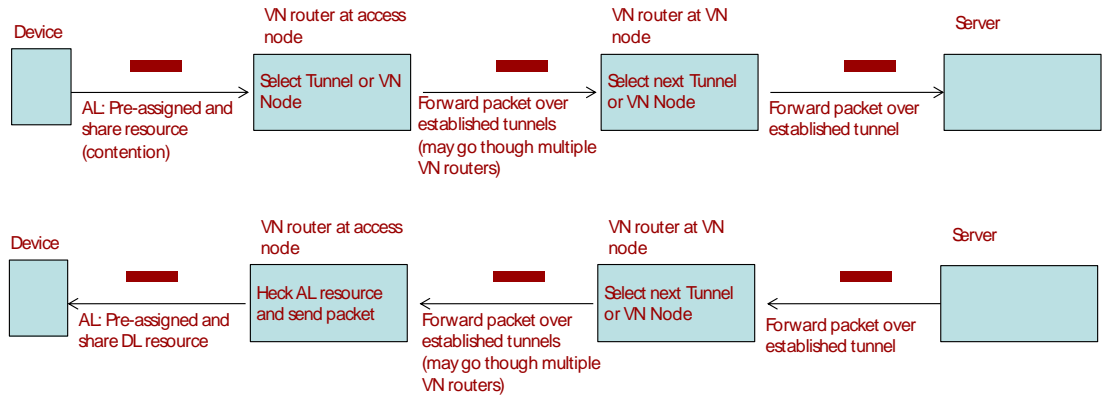

(a). End-to-end packets delivery – fixed end-point case.

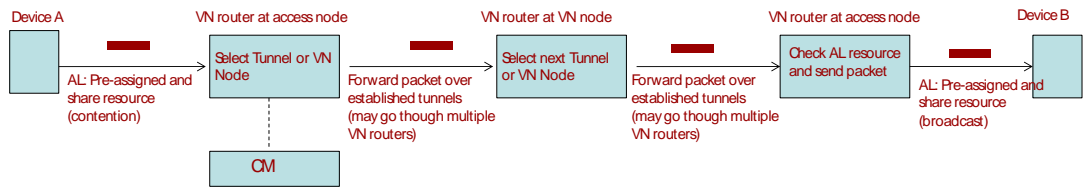

(b) End-to-end packets delivery – mobile end-point case.

Figure 8. End-to-end data packet delivery over a VN using hop-on..

All signaling related to session establishment, re-establishment for a device (session) in 4G can be removed. Data packet transmission from end-points become simply sending data to network using well established VN resource. This procedure can be a completely signaling-free procedure or a lightweight signaling procedure where only AL signaling may be needed if there is no pre-assigned AL resource to a VN or the pre-assigned resource is not shared among end-points of the VN.

## 6. Benefits of slicing and Hop-On (a slice)

### 6.1 What Slice and Hop-on bring to operators?

Slicing concept provides operators tremendous capability to provide right-fit network infrastructure resource to satisfy the huge variety of customer service attributes and quality requirements. SOANC-Com and NFV together provide the definition, realization and policy used by SONAC-Op during VN operation phase without human involvement.  A well defined and realized slice (or VN in the paper) should significantly simplify the operation during slice operation for real-time customer service delivery. Hop-On concept significantly simplifies the data traffic delivery procedure and reduces signaling overhead and latency.

Slicing and Hop-On together enable

- Scalability - network works on service level (integrated data traffic of a service), instead of on device/session level
- Simplicity - Per device/session end-to-end tunnel establishment is removed or minimized
- Flexibility – service customized VN/slice – best fit the need of customers and operators

## 6.2 Use case discussion

The benefits of these concepts are discussed for a few key use cases.

- Massive machine communication – true always-on communication

With predicted millions of millions of machines or device communication, the number of connections has been one of big concerns of operators. With a well defined slice, true always-on of massive machine communication is enabled. Any end-points of a VN designed for such services can simply hop-on the VN by sending data packets using pre-assigned AL resource of this VN. Once data packets get to network, these packets will be routed to the destinations along pre-defined VN tunnels. For VNs where there is no pre-assigned AL resource, only a few signaling message exchanges on AL are needed for data packets transmission over AL. From the point of view of machine or device, the network is always ready to deliver data traffic. The concern of millions of millions of connections can be relieved. A true always-on becomes possible.

- Mobile devices communication and handover-free

With the well defined VN for mobile devices, the data traffic delivery to a mobile device becomes the selection of the right tunnel(s), instead of the re-establishment of new per-device end-to-end connections. One or multiple tunnels at RAN cluster can be selected to enable data delivery by multiple access nodes. True handover-free can be easily enabled by appropriate tunnel selection and rate splitting if needed. The name based location tracking and resolution, and the hierarchical architecture of CM become the one of key enablers of Hop-On and helps the signaling reduction.

- Direct data forwarding

With the well defined VN for peer-to-peer communication (e.g., we-chat), data packets can be routed to the destination via the right tunnels, i.e., shortest route without going through un-necessary paths. CM techniques provide the location information to VN routers to enable the high efficient data delivery.

- Reliable services

A slice can be designed to best fit the customer service requirement. For customer services that require high reliability, a various redundancy techniques can be considered when a VN is designed by SONAC. Redundancy techniques include spatial redundancy, i.e., transmission over multiple tunnels in a RAN cluster, multiple re-transmission techniques which require resource assignment more than source date rate, etc. SONAC will translate the requirement of high reliability into appropriate network resource assignment. During the slice operation, the SONAC-Op will use the assigned resource for this VN to perform data traffic delivery based on the policy configured by SONAC-Com. All end-points of this VN simply send data packets to network. The required reliability will be automatically enabled.

- Low end-to-end latency

For customer services with low end-to-end latency requirement, a VN can be designed with dedicated resource assignment for part or all logical tunnels. The cloud resource can also be dedicated for this VN. By this design, the low E2E latency can be well managed.

- Low response latency service

Some customer services require low response latency at application layer. SONAC can design a VN with customer application functions (servers) implemented at the edge of network, i.e., at RAN cluster and even in access nodes. The logical topology of such a VN includes customer service functions distributed over edge of a network. After the association between devices and in-network application functions, the data traffic can be routed along the pre-defined tunnels. With such a well defined VN, the interaction between devices and application servers can be reduced.

## 7. Summary

The discussed concept of slicing and hop-on (a slice) is one of approaches to address challenges in future networks. This approach is expected to suitable and applicable to most of future use cases. For device (customer) with extreme critical service requirement, per device (customer) slice (VN) can be designed. In this case, on-demand VN creation, adaptation and device hop-on need to be considered jointly.

For sake of simplified description, VN, instead of slice, is used in this paper and one VN serving one service is assumed. If a VN serves multiple customer services, VN ID in this paper can be replaced by VN ID and service ID.

In summary, the concepts of slicing and hop-on (a slice) are expected to bring revolutionary transform to future wireless network. The main challenges, however, are SONAC-Com, which requires intelligent strategy to create multiple VNs and continuously maintain these VNs; and SONAC-Op, which requires intelligent strategy to most efficiently use these VNs for end-point traffic delivery over VNs. To deal with these challenges will be continuously the focus of future research for future network architecture design.

## 8. Reference

[1]     NGMN 5G Project Requirements & Architecture –Work Stream E2E Architecture, Description of Network Slicing Concept Version 1.0, 13th January 2016, by NGMN Alliance

[2]     NGMN 5G White Paper – Executive Version, Version 1.0, December 22,2014: http://www.ngmn.org/uploads/media/141222_NGMNExecutive_Version_of_the_5G_White_Paper_v1_0_01.pdf

[3]     3GPP TR 23.799, Study on Architecture for Next Generation System, version 0.7.0, August, 05, 2016

[4] Hang Zhang, et al, "Future wireless network: MyNET and SONAC", *Network Magazine, IEEE*, Volume: 29, Issue: 4, Page(s): 14 – 23, July-August 2015

[5] Original NFV White Paper, October 2012: http://portal.etsi.org/NFV/NFV_White_Paper.pdf

[6] NFV White paper #2, October 15-17, 2013: http://portal.etsi.org/NFV/NFV_White_Paper2.pdf

[7] NFV White Paper #3 October 14-17, 2014: http://portal.etsi.org/NFV/NVF White Paper3.pdf

[8] ONF White Paper, Software-Defined Networking: The New Norm for Networks, April 13, 2012: https://www.opennetworking.org/images/stories/downloads/sdn-resources/white-papers/wp-sdn-newnorm.pdf

[9] OpenFlow White Paper: Enabling Innovation in Campus Networks, March 14, 2008: http://archive.openflow.org/documents/openflow-wp-latest.pdf

[10] Open Flow-enabled SDN and Network Functions Virtualization, February 17, 2014: https://www.opennetworking.org/images/stories/downloads/sdn-resources/solution-briefs/sb-sdn-nvf-solution.pdf